\documentclass[traditabstract]{aa} 
\usepackage{graphicx}
\usepackage{txfonts}
\usepackage{natbib}
\bibpunct{(}{)}{;}{a}{}{,} 

\begin{document}
   \title{Constraints on photoevaporation models from (lack of) radio emission in the Corona Australis protoplanetary disks.}


   \author{Roberto Galv\'an-Madrid\inst{1,2}
         \and
         Hauyu Baobab Liu\inst{3}
         \and
         Carlo Felice Manara\inst{1}
         \and
	 Jan Forbrich\inst{4}
	 \and 
         Ilaria Pascucci\inst{5}
	 \and 
	 Carlos Carrasco-Gonz\'alez\inst{2}
	 \and 
	 Ciriaco Goddi\inst{6}
	 \and
	 Yasuhiro Hasegawa\inst{3}
         \and  
	 Michihiro Takami\inst{3}
         \and 
	 Leonardo Testi\inst{1,7,8}
          }

\institute{
  European Southern Observatory, Karl-Schwarzschild-Str. 2, 85748 Garching, Germany
  \and 
  Centro de Radioastronom\'ia y Astrof\'isica, Universidad Nacional Aut\'onoma de M\'exico, Morelia 58090, Mexico
  \email{r.galvan@crya.unam.mx}  
  \and 
  Academia Sinica Institute of Astronomy and Astrophysics, P.O. Box 23-141, Taipei, 106, Taiwan
  \and 
  Department of Astrophysics, University of Vienna, T\"urkenschanzstra\ss e 17, A-1180, Vienna, Austria
  \and 
  Lunar and Planetary Laboratory, University of Arizona, Tucson, AZ 85721, USA
  \and
  Joint Institute for VLBI in Europe, Postbus 2, NL-7990 AA Dwingeloo, the Netherlands
  \and 
  INAF-Osservatorio Astrofisico di Arcetri, Largo E. Fermi, I-50125 Firenze, Italy 
  \and 
  Excellence Cluster Universe, Boltzmannstr. 2, D-85748 Garching, Germany 
  }

   \date{Received 18 July, 2014; accepted ?? ??, ????}

 
  \abstract
{
Photoevaporation due to high-energy stellar photons is thought to be one of the main drivers of 
protoplanetary disk dispersal. 
The fully or partially ionized disk surface is expected to produce free-free continuum 
emission at centimeter (cm) wavelengths 
that can be routinely detected with interferometers such as the upgraded Very Large Array (VLA).  
We use deep (rms noise down to 8 $\mu$Jy beam$^{-1}$ in the field of view center) 3.5 cm maps of the nearby (130 pc) 
Corona Australis (CrA) star formation (SF) region to constrain 
disk photoevaporation models. We find that the radio emission from disk sources in CrA is surprisingly 
faint. Only 3 out of 10 sources within the field of view are detected, with flux densities 
of order $10^2$ $\mu$Jy. However, a significant fraction of their emission is non-thermal. 
Typical upper limits for non-detections are $3\sigma\sim 60~\mu$Jy 
beam$^{-1}$.  
Assuming analytic expressions for the free-free emission from extreme-UV (EUV) irradiation,
we derive stringent upper limits to 
the ionizing photon luminosity impinging on the disk surface 
$\Phi_\mathrm{EUV}<1-4\times10^{41}$ s$^{-1}$. 
These  
limits constrain $\Phi_\mathrm{EUV}$ to the low end of the values 
needed by EUV-driven photoevaporation models to clear protoplanetary disks in the 
observed few Myr timescale.
Therefore, at least in CrA, 
EUV-driven photoevaporation is not likely to be the main agent of disk dispersal.
We also compare the observed X-ray luminosities $L_X$ of disk sources with models 
in which photoevaporation is driven by such photons. Although predictions are less specific 
than for the EUV case, most of the observed fluxes (upper limits) are roughly consistent with 
the (scaled) predictions. 
Deeper observations, as well as  predictions spanning a wider parameter space,  
are needed to properly test X-ray driven photoevaporation.}  

  \authorrunning{Galv\'an-Madrid et al.}
  \titlerunning{Constraints on disk photoevaporation models}

   \maketitle
%

\section{Introduction}

The mechanisms that drive the dispersal of disks around young stars are not well understood. 
Photoevaporation of the disk driven by  high-energy radiation from the central star is 
thought to act in concert with viscous accretion and the formation of planets. 
A recent review of the topic is presented by \cite{Alexander13}. 
Several models have been put forward to describe disk photoevaporation, 
including  analytical flow solutions \citep[e.g.,][]{Hollenbach94,Gorti09a} 
and hydrodynamical simulations \citep[e.g.,][]{Alexander06,Owen10}. 
Mid-infrared forbidden lines such as the [Ne II] have 
been used to infer the presence of photoevaporative flows \citep[e.g.,][]{PS09} 
from disks around low-mass young stellar objects (YSOs).  

\cite{Pascucci12} and \cite{Owen13} (hereafter Pascucci12 and Owen13, respectively) 
published predictions for the radio-continuum emission from (partially) ionized disk surfaces 
around low-mass YSOs that may be photoevaporating. 
Pascucci12 showed that free-free emission from the disk surface is 
directly proportional to the ionizing stellar 
radiation\footnote{The extreme-UV (EUV) radiation, with $13.6 < h\nu < 100$ eV.} 
reaching the disk. 
If the photoevaporation is driven by X-rays ($h\nu > 0.1$ keV), 
the interaction of photons with matter is more complex and the 
resulting gas temperatures and ionization fraction can depart considerably from the 
$\sim 10^4$ K and $\sim 1$ characteristic of the EUV case.
The effects of far-UV radiation ($6 < h\nu < 13.6$ eV) add further chemical complexity 
\citep{Gorti09a},  
but this type of radiation is relevant even in photoevaporation models dominated by X-ray 
heating since it regulates the destruction of molecular coolants. 
Deep radio observations of many protoplanetary disks are needed to better constrain 
photoevaporation models. 
Recently, \cite{Pascucci14} analyzed the cm emission from 14 circumstellar disks and found EUV 
photon luminosities ($\Phi_\mathrm{EUV}$) 
lower than $1 \times 10^{42}$ s$^{-1}$ for sources 
with no jets and lower than $5 \times 10^{40}$ s$^{-1}$ for three older systems in their sample, 
thus placing a tight constraint on photoevaporation models.

In this Letter we analyze the 8.5 GHz (3.5 cm) 
continuum emission of the 10 protoplanetary disks within our VLA field of view 
towards the central part 
of the Corona Australis (hereafter CrA) star formation region. This nearby 
region has been studied in the past at all wavelengths from the cm radio to the 
X-rays 
\citep[e.g.,][and references therein]{Choi08,Lindberg14,Peterson11,Forbrich07}. 
The distance to CrA is $\approx 130\pm10$ pc \citep{NF08}.


\section{Data}

The data analyzed here were presented in \cite{Liu14} (hereafter Liu14) 
and are part of a larger 
VLA\footnote{The National Radio Astronomy Observatory is operated by
Associated Universities, Inc. under cooperative agreement with the
National Science Foundation.} 
program to monitor 
the time variability of radio sources in nearby, low-mass star formation regions. 
One of the main products of the program 
are radio continuum images with sensitivity down to a few microjanskys.

The main continuum map used here concatenates 14 VLA epochs in 2012. It 
covers the continuous frequency range from 8 GHz (3.7 cm) to 9 GHz (3.3 cm) 
and reaches an rms noise of $\sigma\sim8~\mu$Jy beam$^{-1}$ in the center of the field. 
The half-power beam width (HPBW) of the primary beam is 315\arcsec, 
and the beamwidth at 10\% power is 530\arcsec. 
The synthesized HPBW is $4.6\arcsec \times2.1\arcsec$, PA$=-179.4^\circ$. 
Further details of the observations are described in Liu14. 
Analysis of the data was performed in CASA \citep{McMullin07}. 

\section{Results}

\subsection{Source list}

\begin{table*}[]
\caption{\label{t1} Disks in CrA within the VLA primary beam}
\centering
\begin{tabular}{cccccccccc}
\hline
(1) & (2) & (3) & (4) & (5) & (6) & (7) & (8) & (9) & (10) \\
Name   &   RA  &  Dec &  CSA11 &  SA13 &  $S_\mathrm{3.5cm}$ &  
PB-corrected noise  &  $L_X$ & Spectral & Comments \\
       &  [h:m:s] & [deg:arcmin:arcsec] &  & & [$\mu$Jy] & [$\mu$Jy] & [erg s$^{-1}$] & type & \\
\hline
G-95   &  19:01:28.72 & -36:59:31.7 & TD & ... & $<180$ & 60 & $1.03\times10^{30}$ & M1 &  ... \\
G-87   &  19:01:32.32 & -36:58:03.0 & TD & TD & $<60$ & 20 & ... & M1.5 &  ...  \\
G-85 & 19:01:33.85 & -36:57:44.8 & PD  & pre-TD & $<60$ & 20 & $4.28\times10^{29}$ & M0.5 & ...  \\ 
V709   &  19:01:34.84 & -37:00:56.7 & TD/DD & ...  &  $892\pm79$ & 50 & $4.17\times10^{30}$ & K1 & a \\
HBC-677  & 19:01:41.62 & -36:59:53.1 & PD & PD &  $<45$ & 15 & $2.21\times10^{29}$ & M2 & ... \\
IRS8 & 19:01:51.11 & -36:54:12.5 & PD  &  ... &   $<120$ & 40 & ... & M2 & ... \\
R CrA & 19:01:53.67 & -36:57:08.3 & PD  & ... & $285\pm10$ & 10 & $7.7\times10^{29}$ & A5 & b \\
CrA-465 & 19:01:53.74 & -37:00:33.9 & PD & ... &  $<60$ & 20 & ...  & M5-M7.5 &  ... \\
G-32 & 19:01:58.33 & -37:00:26.7 &  PD &  ...  &$<60$ & 20 & ...  & $>\mathrm{M5}$ & ... \\
T CrA & 19:01:58.79 & -36:57:50.1 & PD & PD & $181\pm9$ & 15 &  ... & F0 & c \\
\hline
\end{tabular}
\tablefoot{
Columns 1 to 3: names and coordinates from CSA11 \citep[who use the {\it Chandra} positions of][]{FP07} 
and SA13. Columns 4 and 5: classification in CSA11 and SA13, respectively. Primordial disk (PD), 
pre-transitional disk (pre-TD), transitional disk (TD), debris disk (DD). A blank field means 
that the source is either not detected, or not mentioned, or the photometry was affected by extended 
nebulosity, mostly an issue in the {\it Herschel} observations (SA13). 
Column 6: flux density at 3.5 cm or $3\sigma$ upper limits. For the detections, the VLA coordinates 
from gaussian fits are given as a comment, together with the estimation of the 3.5 cm flux due to 
free-free from the coadded non-detection epochs. 
Column 7: primary-beam corrected noise, measured individually around the position of each source. 
Column 8: X-ray luminosity in the 0.2 -- 8 keV band from \cite{FP07}. A blank field means that only $\sim 10^1$ counts were 
detected, and the X-ray spectral shape and luminosity could not be derived. However, under 
typical assumptions, their measured count number would imply $L_X \sim 1\times10^{29}$ erg s$^{-1}$. 
Column 9: spectral types reported by CSA11. Compiled from \cite{FP07}, \cite{MW09},  
\cite{LopezMarti05}, and CSA11. 
Column 10: comments. a) RA=19:01:34.86, Dec=-37:00.55.8. $S_\mathrm{V709,ff} \lesssim 507\pm88$ $\mu$Jy. 
b) RA=19:01:53.68, Dec=-36:57:08.0. $S_\mathrm{RCrA,ff} \lesssim 183\pm20$ $\mu$Jy. Intermediate-mass YSO  
\citep{Acke04}. 
c) RA=19:01:58.79, Dec=-36:57:49.9. $S_\mathrm{TCrA,ff} \lesssim 170\pm26$ $\mu$Jy. Intermediate-mass YSO 
\citep{Acke04}. 
}
\end{table*}

We compiled a list of the protoplanetary disks in CrA and searched for their radio continuum emission 
in our deep VLA images, restricting ourselves to sources within the 10\% response level of 
the primary beam (the interferometric field of view). 
Protoplanetary disks were identified based on the catalogues of \cite{CurrieSA11} (hereafter CSA11) and 
\cite{SA13} (hereafter SA13). 
CSA11 use {\it Spitzer} IRAC and MIPS photometry and IRS spectroscopy in their models, whereas  
SA13 complement with {\it Herschel} far-infrared PACS photometry.

Table 1 lists the basic properties of the disks in CrA, and figure 1 shows the deep radio 
map with the targets labeled. 
We only consider objects classified as primordial, pre-transitional, or transitional disks 
(roughly equivalent to class II YSOs). 
The age of the considered sources is 1 to 3 Myr (CSA11).

\subsection{The radio-continuum of disks in CrA}

\begin{figure*}
\centering
\includegraphics[scale=0.55]{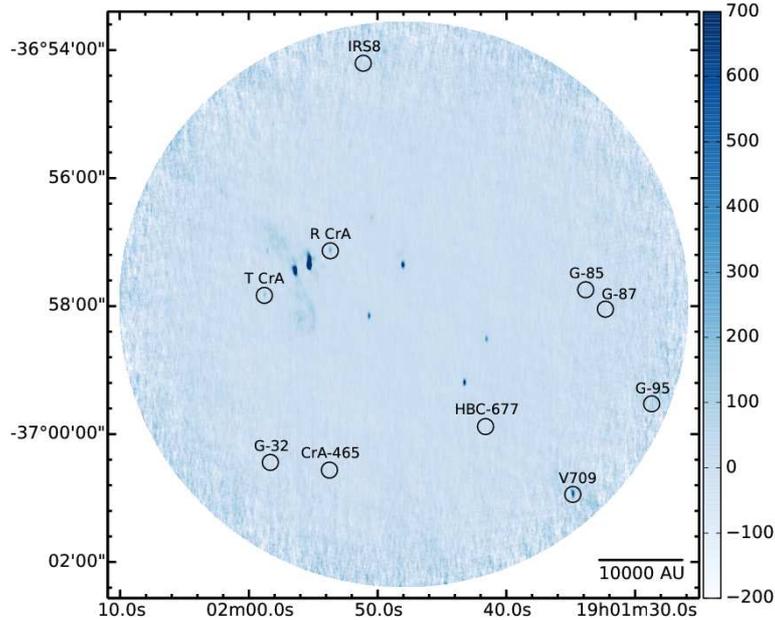} 
\caption{Deep coadded VLA 3.5 cm (8.5 GHz) image of CrA. The image is corrected for primary-beam 
attenuation 
down to 10\% response (530\arcsec diameter). 
The synthesized HPBW is $4.6\arcsec \times2.1\arcsec$, PA$=-179.4^\circ$.  
The disk sources listed in Table 1 are marked with circles and labeled. 
The intensity scale is in units of $\mu$Jy beam$^{-1}$.
}
\label{f1_continuum}
\end{figure*}
%

The radio-continuum emission of YSOs associated with disks can in principle have contributions 
from several physical 
mechanisms. Disk-photoevaporation models emphasize the contribution from the free-free emission 
of the photoionized disk wind (\cite{AvalosLizano12}, Pascucci12, Owen13). 
Free-free emission from magnetohydrodynamical jets dominates in class 0 and I YSOs 
\citep[e.g.,][]{Anglada98,Reipurth04}, but their contribution in class IIs should 
be smaller. 
The models of \cite{Shang04} (aimed at class I YSOs) predict a 3.5 cm flux of 16 $\mu$Jy at 140 pc 
for their lowest jet mass-loss rate calculation of 
$3\times10^{-8}~M_\odot$ yr$^{-1}$. Typical mass-loss rates in class II YSOs are smaller. 
Therefore, we expect the jet free-free contamination to be below the noise. 
In contrast, gyrosynchrotron emission from active magnetospheres can contribute significantly and 
even dominate the radio-continuum emission in class II and III YSOs \citep[e.g.,][]{Gibb99,Forbrich07}. 
Finally, it is also possible that large dust grains emit significantly at short cm wavelenths, 
for example in TW Hydrae (\cite{Wilner05}, Pascucci12). 
However, this may not be typical. 
\cite{Rodmann06} investigated the dust and free-free contributions at 7 mm in protoplanetary 
disks and find that $\sim 80\%$ of the emission is from dust. Assuming a dust spectral index typical 
of disks $\alpha = 2$ to 2.5 \citep[e.g.,][]{Ricci10}, 
this translates into only $\sim 11\%$ to $5\%$ dust emission at 3.5 cm.

We searched for radio continuum emission in our deep map toward the disks listed in Table 1. 
The main result of our search is that 7 out of 10 targets are not detected. 
5 out of the 7 non-detections have $3\sigma$ upper limits 
to the peak intensity $\leq 60$ $\mu$Jy beam$^{-1}$ (see Table 1). 
This sets a tight constraint on disk-photoevaporation models (Section 4).
For the detections, the flux density was measured using the task 
{\it imfit} in CASA. 
Figure A.1 in the online appendix shows the field around each YSO individually.

Although it is difficult to quantify the relative contributions of thermal (free-free) and non-thermal 
(gyrosynchrotron) emission for the detections, the dominant mechanism can be identified from the 
spectral index and the variability properties. 
There is currently no information on the spectral 
index of the radio detections. 
However, the three detected disk sources (V709, R CrA, T CrA) were identified by Liu14 as highly 
variable.
In maps of the 14 individual epochs that were 
coadded in the final deep image, these sources can sometimes  
be detected at a level of a few $\times 10^2$ $\mu$Jy (or up to $\sim 1$ mJy for V709) 
and then be non-detections even between observations separated by as little as $\sim 1$ day 
(Liu14). 
This large variability shows that most of the cm flux of the detected 
sources is not due to stable, free-free emission from a photoevaporative flow, but most likely 
due to non-thermal (gyro)synchrotron emission from magnetic activity closer to the stellar 
surface (e.g., Liu14, Forbrich et al. 2007). 
We estimate the free-free contribution at 3.5 cm in the detections 
by assuming that free-free is relatively stable in time 
compared to (gyro)synchrotron emission (Liu14). Then we make images that 
include the concatenated data from each of the individual epochs in which each of them is not detected, 
and measure the fluxes.  
We obtain free-free fluxes $S_\mathrm{ff,V709} \lesssim 507\pm88~\mu$Jy 
($\sim 57\%$ of the total flux), 
$S_\mathrm{ff,RCrA} \lesssim 183\pm20~\mu$Jy ($\sim 64\%$), and 
$S_\mathrm{ff,TCrA} \lesssim 170\pm26~\mu$Jy ($\sim 94\%$). 
We note that these are still strict upper limits to the free-free emission. 
Also, if any of the extra processes mentioned above is  
significant, it would push down the upper limits to the ionizing fluxes derived in Section 4.

The X-ray properties of the radio-detected sources also hint that radio emission, when bright, 
is mainly due to magnetospheric activity. 
\cite{FP07} presented a deep {\it Chandra} observation of CrA. 
From the spectral shape and time-domain properties of the sources in CrA, those authors conclude 
that the YSOs X-ray emission is dominated by coronal activity, 
consistent with our interpretation of the cm detections as having significant 
coronal (gyro)synchrotron emission. 
Table 1 also lists the X-ray luminosities in the 0.2 -- 8 keV band of the disk sources 
in CrA. Note that 
V709 is the most luminous in X-rays 
\citep[it also shows variations up to $\sim 50\%$ within days,][]{FP07}.  
R CrA is the third most luminous X-ray source and shows   
variations of a factor of 5 within days. In contrast, T CrA is barely detected in X-rays,  
and there is no information on its variability \citep{FP07}.

\section{Implications for models of disk photoevaporation}

We use our observations to constrain possible models of protoplanetary disk photoevaporation. 
First we compare quantitatively with predictions from EUV-driven 
photoevaporation models. 
Because the free-free emission is directly proportional to the EUV luminosity 
 reaching the disk ($\Phi_\mathrm{EUV}$), we can use equation 2 of Pascucci12 to derive 
upper limits to $\Phi_\mathrm{EUV}$.  
The main assumptions here are that the surface of the disk is almost fully ionized at $10^4$ K 
and that the fraction of EUV photons absorbed by the disk is 0.7 \citep[][]{HG09}. 
We then compare with X-ray driven models, for which 
there are still important uncertainties on the assumptions \citep{Alexander13}.  
Therefore, in this case the comparison is more qualitative. 
 
Table 2 summarizes the comparison to the predictions of Pascucci12. 
The constraints to models of EUV-driven photoevaporation are strong. 9 out of 10 
$3\sigma$  
upper limits to the ionizing-photon luminosity are in the range 
$\Phi_\mathrm{EUV} < 1-4 \times 10^{41}$ s$^{-1}$.
The EUV photon luminosity emitted by young pre-main sequence stars is constrained only 
for a handful of sources but is always found to be $>10^{41}$ and often 
$>10^{42}$ s$^{-1}$ \citep{Alexander05,Herczeg07}. 
However, these results are dependent on reddening, and not all the emitted photons 
necessarily reach the disk.
Theoretical models typically use $\Phi_\mathrm{EUV}=10^{41}$ s$^{-1}$ or higher. 
On the lower end, \cite{Font04} calculates models for $\Phi_\mathrm{EUV}=10^{40}$ to $10^{42}$ s$^{-1}$. 
More recent models that put emphasis in satisfying the constraint of a disk lifetime of a few Myr 
\citep[e.g.,][]{Hernandez07} 
use $\Phi_\mathrm{EUV}=10^{42}$ s$^{-1}$ \citep[e.g.,][]{Alexander06}.  
\cite{AlexanderArmitage09} present a grid of models with coupled photoevaporation, 
viscous transport, 
and Type II migration. Using $\Phi_\mathrm{EUV}=10^{42}$ s$^{-1}$, those 
authors find a median disk lifetime of 4 Myr, consistent with observational constraints. 
Our observations show that, at least for most of the disks in CrA, $\Phi_\mathrm{EUV}$ 
is on the low side of the broad range of $\Phi_\mathrm{EUV}$ used in models.

Upper limits to $L_X$ from Pascucci12 are listed in Table 2. 
These limits are at least one order of magnitude higher than the observed $L_X$ (Table 1). 
Therefore, if $S_\mathrm{ff} \propto L_X$, from these models 
we would expect to detect the sources at a level of $\sim 1$ to 10 $\mu$Jy 
(below our sensitivity). 
For the 3 detections, the observed cm flux appears to be too high with respect to 
the model predictions. This could be explained if the stable cm flux, which we tentatively 
attribute to free-free emission from photoevaporation, 
still has contributions from other processes (Section 3.2).
 
We now compare to the 
radiative transfer calculations on hydrodynamical simulations presented by Owen13, who use  
a fiducial X-ray luminosity $L_X = 2 \times 10^{30}$ erg s$^{-1}$. 
We note that for a given $L_X$, Owen13 predicts a larger free-free flux than 
Pascucci12. 
The difference could be due to Owen13 using an unattenuated stellar spectrum, which may still 
contain substantial EUV and soft-Xray emission, but this is not certain. 
If we scale the frequency and distance of their predictions to 8 GHz and 130 
pc\footnote{Owen13 provides spectral 
indices between 8 GHz and their anchor predictions at 15 GHz of $\alpha \sim 0.5$ for 
X-ray driven photoevaporation.}, we find that the expected 3.5 cm flux is 
$S \sim 2 \times 10^2$ $\mu$Jy (see figure 11 in Owen13). 
From Table 1 we see that, roughly, the 3 sources with $L_X$ close to $10^{30}$ erg s$^{-1}$ have 
free-free fluxes (V709\footnote{Note that the evolutionary stage of V709 is not clear: it is the 
only source in the sample that could be a debris disk, 
in which case no photoevaporation signal is expected.}, 
R CrA) or upper limits (G-95) consistent with these model predictions. 
We do not compare the results of Owen13 with the rest of our sources with lower $L_X$, 
since it is not clear which fraction of the predicted free-free emission is only due to 
the X-rays in their input stellar spectrum.
 
Since photoevaporation driven purely by X-rays would produce lower free-free fluxes, 
deeper observations are needed to probe this regime.
Model predictions spanning parameter space towards lower EUV and X-ray luminosities 
and different X-ray hardness are also needed.

\begin{table}[t]
\caption{\label{t1} Comparison to Pascucci12.}
\centering
\begin{tabular}{ccc}
\hline
Name & $\Phi_\mathrm{EUV,model}$ & $L_\mathrm{X,model}$  \\
     & [s$^{-1}$]      &  [erg s$^{-1}$] \\
\hline
G-95  &  $< 4.0\times10^{41}$  &  $< 4.9\times10^{31}$ \\
G-87  &  $< 1.3\times10^{41}$  &  $< 1.6\times10^{31}$ \\
G-85  &  $< 1.3\times10^{41}$  & $< 1.6\times10^{31}$ \\
V709  &  $\lesssim 1.1\times10^{42}$ &  $\lesssim 1.4\times10^{32}$ \\
HBC-677 & $< 1.0\times10^{41}$ & $< 1.2\times10^{31}$ \\
IRS 8  & $< 2.7\times10^{41}$  & $< 3.2\times10^{31}$ \\
R CrA  & $\lesssim 4.1\times10^{41}$ & $\lesssim 5.0\times10^{31}$ \\
CrA-465 & $< 1.3\times10^{41}$  &  $< 1.6\times10^{31}$ \\
G-32  & $< 1.3\times10^{41}$  &  $< 1.6\times10^{31}$ \\
T CrA & $\lesssim 3.8\times10^{41}$ & $\lesssim 4.6\times10^{31}$ \\
\hline
\end{tabular}
\tablefoot{
Upper limits using equations 2 and 3 of Pascucci12. The 7 non-detections are marked 
as simple upper limits with a $<$ symbol. For the 3 detections, we use the estimation of their 
free-free flux from the coadded non-detections, which is still an strict upper limit to the free-free 
emission from a photoevaporating disk (see Section 2). We mark them with a $\lesssim$ 
symbol.  
}
\end{table}

\section{Conclusions}

With the purpose of constraining models of protoplanetary disk clearing via 
photoevaporation, we inspect the 8.5 GHz (3.5 cm) continuum emission of disks in the nearby (130 pc) 
CrA star formation region. We use the deep (noise down to 8 $\mu$Jy 
beam$^{-1}$ in the center of the field of view) maps from our monitoring survey of low mass YSOs (Liu14). 

We find that disks are radio faint: 7 out of 10 targets are not detected. 
5 of the non-detections have $3\sigma$ upper limits $\leq 60~\mu$Jy beam$^{-1}$, and all of 
them have $3\sigma \leq 180~\mu$Jy beam$^{-1}$.
Furthermore, for the 3 radio detections, their 
radio variability and X-ray 
properties indicate that a significant fraction of their radio flux is due to non-thermal processes, 
rather than due to the putative photoevaporative flow. 
We measure the stable 3.5 cm flux ($\sim 170$ to 507 $\mu$Jy) 
of the 3 detections and attribute it to free-free emission. Strictly, these are still upper limits 
to the fluxes from photoevaporation.

Using the prescription of Pascucci12 for EUV-driven photoevaporation, we derive tight upper 
limits to the rate of ionizing photons reaching the disk 
$\Phi_\mathrm{EUV}$:  
9 out of 10 disks in our field of view have 
$\Phi_\mathrm{EUV}<1-4\times10^{41}$ s$^{-1}$. 
These upper limits  
discard the higher end
of what EUV-driven photoevaporation models require to clear a protoplanetary disk in the observed 
timescale of a few Myr. 
Also, the limits we derive from radio observations are lower than the previous estimations  
of the EUV photons emitted by the star 
derived by \cite{Alexander05} and \cite{Herczeg07}.  
This suggests that a significant fraction of the emitted EUV 
photons does not reach the disk. 
Our results, together with those recently reported by \cite{Pascucci14}, show 
that EUV photons are unlikely to be the main driver in disk dispersal.  

We also compare to models of X-ray driven photoevaporation. This comparison is more 
qualitative because there is more room for variation in the model assumptions. 
Most of the observed fluxes (upper limits) are roughly consistent with 
the (scaled) predictions. Some detections appear to be too bright in the radio. However, 
their stable cm flux could still have some contamination.

Future, deeper maps of this and other star formation regions (including variability information), 
together with  X-ray and near-infrared data (to obtain accretion rates) will help to set 
tighter constraints on the mechanisms that drive protoplanetary disk photoevaporation.  
More specific predictions, in particular for lower EUV and X-ray luminosities, are also needed.

\begin{acknowledgements}
This research made use of APLpy, an open-source plotting package for Python hosted at 
http://aplpy.github.com. 
R.G.-M. acknowledges funding from the European Community's Seventh 
Framework Programme (/FP7/2007-2013/) under grant agreement No. 229517R.
I.P. acknowledges support from the NSF Astronomy \& Astrophysics Research Grant 1312962.
The authors thank the anonymous referee for an insightful report. 
\end{acknowledgements}








   
  


\bibliographystyle{aa} 
\bibliography{references} 

\begin{thebibliography}{36}
\expandafter\ifx\csname natexlab\endcsname\relax\def\natexlab#1{#1}\fi

\bibitem[{{Acke} \& {van den Ancker}(2004)}]{Acke04}
{Acke}, B. \& {van den Ancker}, M.~E. 2004, \aap, 426, 151

\bibitem[{{Alexander} {et~al.}(2013){Alexander}, {Pascucci}, {Andrews},
  {Armitage}, \& {Cieza}}]{Alexander13}
{Alexander}, R., {Pascucci}, I., {Andrews}, S., {Armitage}, P., \& {Cieza}, L.
  2013, ArXiv e-prints

\bibitem[{{Alexander} \& {Armitage}(2009)}]{AlexanderArmitage09}
{Alexander}, R.~D. \& {Armitage}, P.~J. 2009, \apj, 704, 989

\bibitem[{{Alexander} {et~al.}(2005){Alexander}, {Clarke}, \&
  {Pringle}}]{Alexander05}
{Alexander}, R.~D., {Clarke}, C.~J., \& {Pringle}, J.~E. 2005, \mnras, 358, 283

\bibitem[{{Alexander} {et~al.}(2006){Alexander}, {Clarke}, \&
  {Pringle}}]{Alexander06}
{Alexander}, R.~D., {Clarke}, C.~J., \& {Pringle}, J.~E. 2006, \mnras, 369, 229

\bibitem[{{Anglada} {et~al.}(1998){Anglada}, {Villuendas}, {Estalella},
  {Beltr{\'a}n}, {Rodr{\'{\i}}guez}, {Torrelles}, \& {Curiel}}]{Anglada98}
{Anglada}, G., {Villuendas}, E., {Estalella}, R., {et~al.} 1998, \aj, 116, 2953

\bibitem[{{Avalos} \& {Lizano}(2012)}]{AvalosLizano12}
{Avalos}, M. \& {Lizano}, S. 2012, \apj, 751, 63

\bibitem[{{Choi} {et~al.}(2008){Choi}, {Hamaguchi}, {Lee}, \&
  {Tatematsu}}]{Choi08}
{Choi}, M., {Hamaguchi}, K., {Lee}, J.-E., \& {Tatematsu}, K. 2008, \apj, 687,
  406

\bibitem[{{Currie} \& {Sicilia-Aguilar}(2011)}]{CurrieSA11}
{Currie}, T. \& {Sicilia-Aguilar}, A. 2011, \apj, 732, 24

\bibitem[{{Font} {et~al.}(2004){Font}, {McCarthy}, {Johnstone}, \&
  {Ballantyne}}]{Font04}
{Font}, A.~S., {McCarthy}, I.~G., {Johnstone}, D., \& {Ballantyne}, D.~R. 2004,
  \apj, 607, 890

\bibitem[{{Forbrich} \& {Preibisch}(2007)}]{FP07}
{Forbrich}, J. \& {Preibisch}, T. 2007, \aap, 475, 959

\bibitem[{{Forbrich} {et~al.}(2007){Forbrich}, {Preibisch}, {Menten},
  {Neuh{\"a}user}, {Walter}, {Tamura}, {Matsunaga}, {Kusakabe}, {Nakajima},
  {Brandeker}, {Fornasier}, {Posselt}, {Tachihara}, \& {Broeg}}]{Forbrich07}
{Forbrich}, J., {Preibisch}, T., {Menten}, K.~M., {et~al.} 2007, \aap, 464,
  1003

\bibitem[{{Gibb}(1999)}]{Gibb99}
{Gibb}, A.~G. 1999, \mnras, 304, 1

\bibitem[{{Gorti} \& {Hollenbach}(2009)}]{Gorti09a}
{Gorti}, U. \& {Hollenbach}, D. 2009, \apj, 690, 1539

\bibitem[{{Herczeg}(2007)}]{Herczeg07}
{Herczeg}, G.~J. 2007, in IAU Symposium, Vol. 243, IAU Symposium, ed.
  J.~{Bouvier} \& I.~{Appenzeller}, 147--154

\bibitem[{{Hern{\'a}ndez} {et~al.}(2007){Hern{\'a}ndez}, {Hartmann}, {Megeath},
  {Gutermuth}, {Muzerolle}, {Calvet}, {Vivas}, {Brice{\~n}o}, {Allen},
  {Stauffer}, {Young}, \& {Fazio}}]{Hernandez07}
{Hern{\'a}ndez}, J., {Hartmann}, L., {Megeath}, T., {et~al.} 2007, \apj, 662,
  1067

\bibitem[{{Hollenbach} \& {Gorti}(2009)}]{HG09}
{Hollenbach}, D. \& {Gorti}, U. 2009, \apj, 703, 1203

\bibitem[{{Hollenbach} {et~al.}(1994){Hollenbach}, {Johnstone}, {Lizano}, \&
  {Shu}}]{Hollenbach94}
{Hollenbach}, D., {Johnstone}, D., {Lizano}, S., \& {Shu}, F. 1994, \apj, 428,
  654

\bibitem[{{Lindberg} {et~al.}(2014){Lindberg}, {J{\o}rgensen}, {Brinch},
  {Haugb{\o}lle}, {Bergin}, {Harsono}, {Persson}, {Visser}, \&
  {Yamamoto}}]{Lindberg14}
{Lindberg}, J.~E., {J{\o}rgensen}, J.~K., {Brinch}, C., {et~al.} 2014, \aap,
  566, A74

\bibitem[{{Liu} {et~al.}(2014){Liu}, {Galv{\'a}n-Madrid}, {Forbrich},
  {Rodr{\'{\i}}guez}, {Takami}, {Costigan}, {Felice Manara}, {Yan}, {Karr},
  {Chou}, {Ho}, \& {Zhang}}]{Liu14}
{Liu}, H.~B., {Galv{\'a}n-Madrid}, R., {Forbrich}, J., {et~al.} 2014, \apj,
  780, 155

\bibitem[{{L{\'o}pez Mart{\'{\i}}} {et~al.}(2005){L{\'o}pez Mart{\'{\i}}},
  {Eisl{\"o}ffel}, \& {Mundt}}]{LopezMarti05}
{L{\'o}pez Mart{\'{\i}}}, B., {Eisl{\"o}ffel}, J., \& {Mundt}, R. 2005, \aap,
  444, 175

\bibitem[{{McMullin} {et~al.}(2007){McMullin}, {Waters}, {Schiebel}, {Young},
  \& {Golap}}]{McMullin07}
{McMullin}, J.~P., {Waters}, B., {Schiebel}, D., {Young}, W., \& {Golap}, K.
  2007, in Astronomical Society of the Pacific Conference Series, Vol. 376,
  Astronomical Data Analysis Software and Systems XVI, ed. R.~A. {Shaw},
  F.~{Hill}, \& D.~J. {Bell}, 127

\bibitem[{{Meyer} \& {Wilking}(2009)}]{MW09}
{Meyer}, M.~R. \& {Wilking}, B.~A. 2009, \pasp, 121, 350

\bibitem[{{Neuh{\"a}user} \& {Forbrich}(2008)}]{NF08}
{Neuh{\"a}user}, R. \& {Forbrich}, J. 2008, {The Corona Australis Star Forming
  Region}, ed. B.~{Reipurth}, 735

\bibitem[{{Owen} {et~al.}(2010){Owen}, {Ercolano}, {Clarke}, \&
  {Alexander}}]{Owen10}
{Owen}, J.~E., {Ercolano}, B., {Clarke}, C.~J., \& {Alexander}, R.~D. 2010,
  \mnras, 401, 1415

\bibitem[{{Owen} {et~al.}(2013){Owen}, {Scaife}, \& {Ercolano}}]{Owen13}
{Owen}, J.~E., {Scaife}, A.~M.~M., \& {Ercolano}, B. 2013, \mnras, 434, 3378

\bibitem[{{Pascucci} {et~al.}(2012){Pascucci}, {Gorti}, \&
  {Hollenbach}}]{Pascucci12}
{Pascucci}, I., {Gorti}, U., \& {Hollenbach}, D. 2012, \apjl, 751, L42

\bibitem[{{Pascucci} {et~al.}(2014){Pascucci}, {Ricci}, {Gorti}, {Hollenbach},
  {Hendler}, {Brooks}, \& {Contreras}}]{Pascucci14}
{Pascucci}, I., {Ricci}, L., {Gorti}, U., {et~al.} 2014, ArXiv e-prints

\bibitem[{{Pascucci} \& {Sterzik}(2009)}]{PS09}
{Pascucci}, I. \& {Sterzik}, M. 2009, \apj, 702, 724

\bibitem[{{Peterson} {et~al.}(2011){Peterson}, {Caratti o Garatti}, {Bourke},
  {Forbrich}, {Gutermuth}, {J{\o}rgensen}, {Allen}, {Patten}, {Dunham},
  {Harvey}, {Mer{\'{\i}}n}, {Chapman}, {Cieza}, {Huard}, {Knez}, {Prager}, \&
  {Evans}}]{Peterson11}
{Peterson}, D.~E., {Caratti o Garatti}, A., {Bourke}, T.~L., {et~al.} 2011,
  \apjs, 194, 43

\bibitem[{{Reipurth} {et~al.}(2004){Reipurth}, {Rodr{\'{\i}}guez}, {Anglada},
  \& {Bally}}]{Reipurth04}
{Reipurth}, B., {Rodr{\'{\i}}guez}, L.~F., {Anglada}, G., \& {Bally}, J. 2004,
  \aj, 127, 1736

\bibitem[{{Ricci} {et~al.}(2010){Ricci}, {Testi}, {Natta}, {Neri}, {Cabrit}, \&
  {Herczeg}}]{Ricci10}
{Ricci}, L., {Testi}, L., {Natta}, A., {et~al.} 2010, \aap, 512, A15

\bibitem[{{Rodmann} {et~al.}(2006){Rodmann}, {Henning}, {Chandler}, {Mundy}, \&
  {Wilner}}]{Rodmann06}
{Rodmann}, J., {Henning}, T., {Chandler}, C.~J., {Mundy}, L.~G., \& {Wilner},
  D.~J. 2006, \aap, 446, 211

\bibitem[{{Shang} {et~al.}(2004){Shang}, {Lizano}, {Glassgold}, \&
  {Shu}}]{Shang04}
{Shang}, H., {Lizano}, S., {Glassgold}, A., \& {Shu}, F. 2004, \apjl, 612, L69

\bibitem[{{Sicilia-Aguilar} {et~al.}(2013){Sicilia-Aguilar}, {Henning}, {Linz},
  {Andr{\'e}}, {Stutz}, {Eiroa}, \& {White}}]{SA13}
{Sicilia-Aguilar}, A., {Henning}, T., {Linz}, H., {et~al.} 2013, \aap, 551, A34

\bibitem[{{Wilner} {et~al.}(2005){Wilner}, {D'Alessio}, {Calvet}, {Claussen},
  \& {Hartmann}}]{Wilner05}
{Wilner}, D.~J., {D'Alessio}, P., {Calvet}, N., {Claussen}, M.~J., \&
  {Hartmann}, L. 2005, \apjl, 626, L109

\end{thebibliography}

\Online

\begin{appendix} 

\section{Individual maps of the targets}

%
\begin{figure}
\centering
\begin{tabular}{ccc}
\includegraphics[scale=0.25]{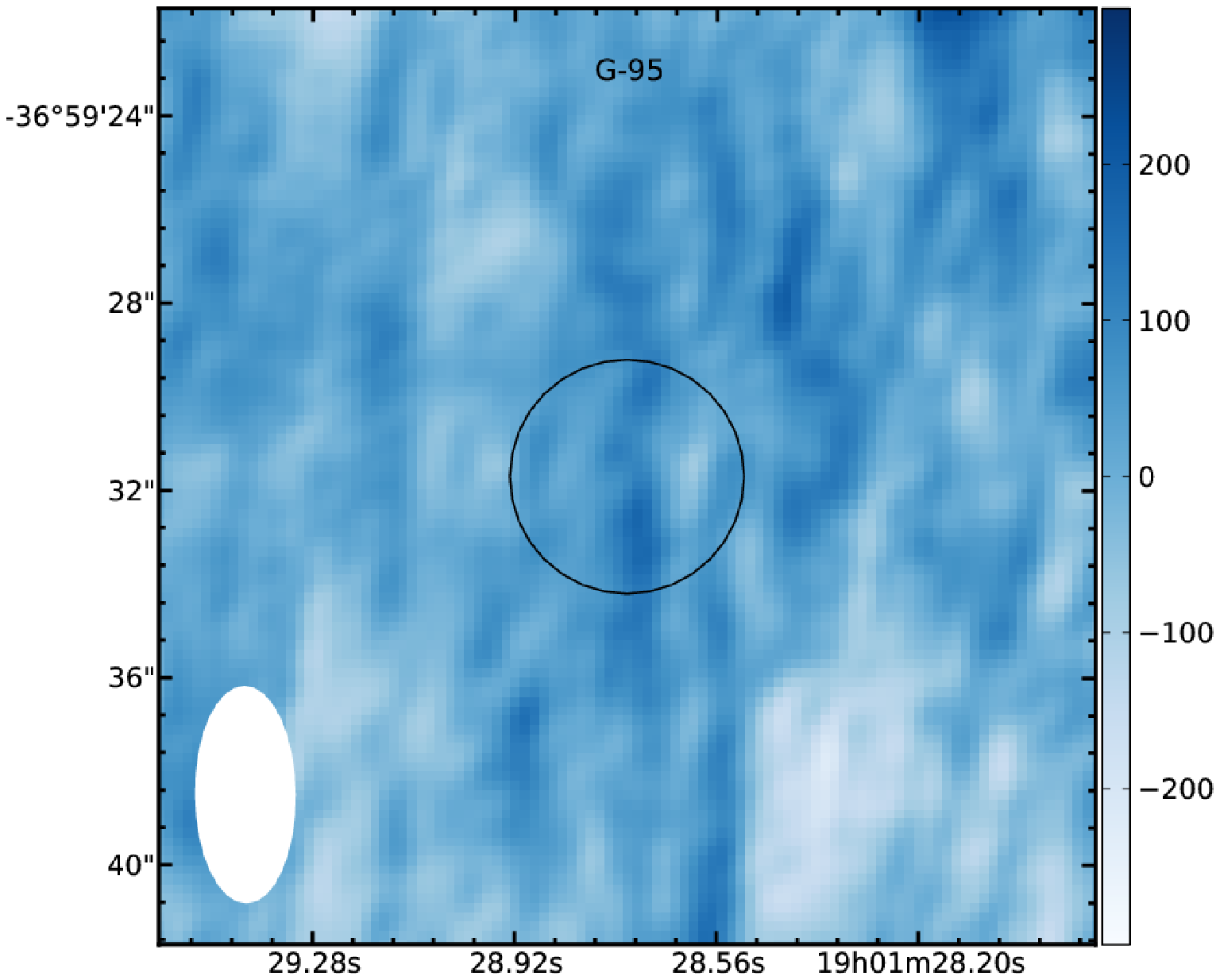} & 
\includegraphics[scale=0.25]{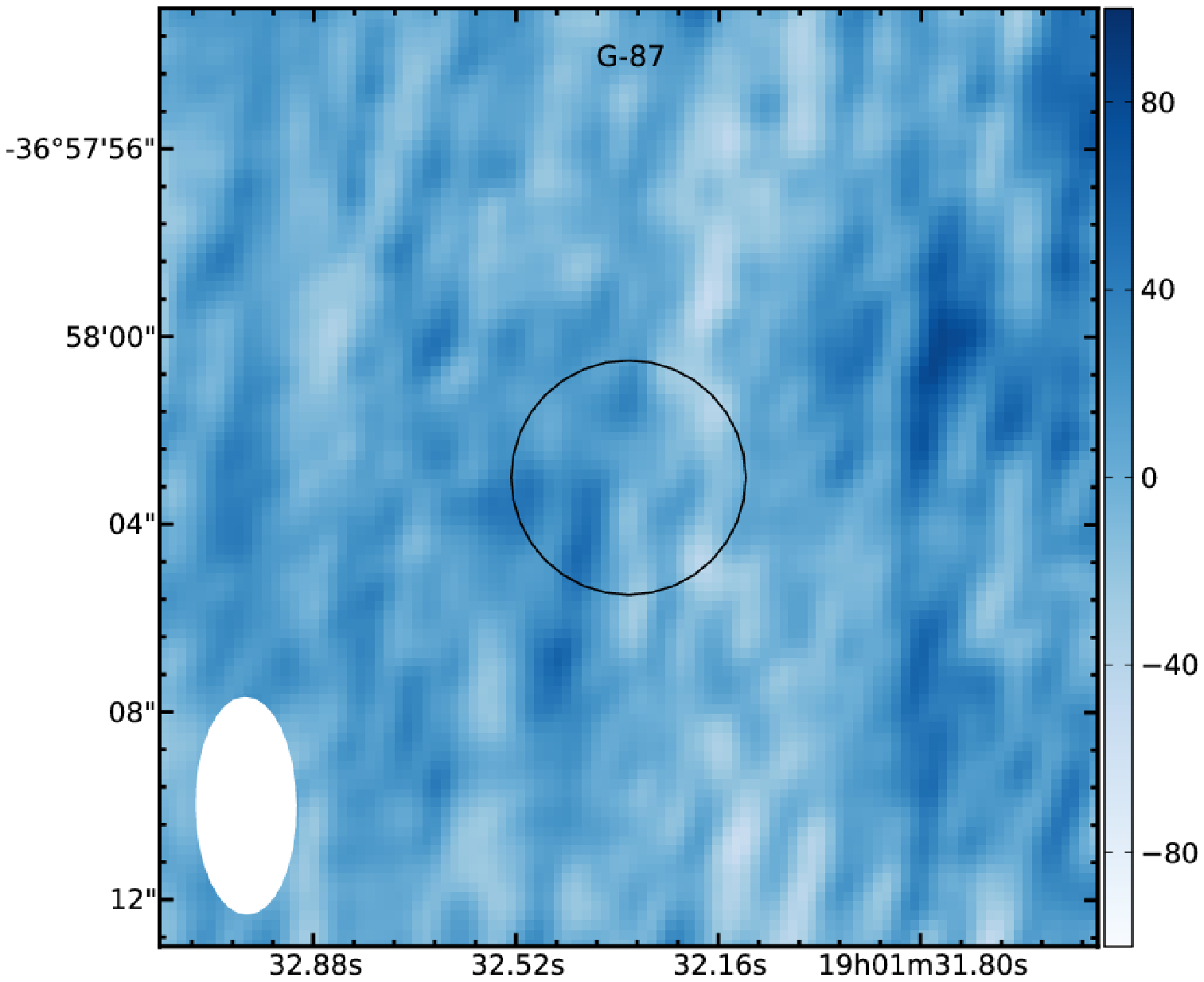} & 
\includegraphics[scale=0.25]{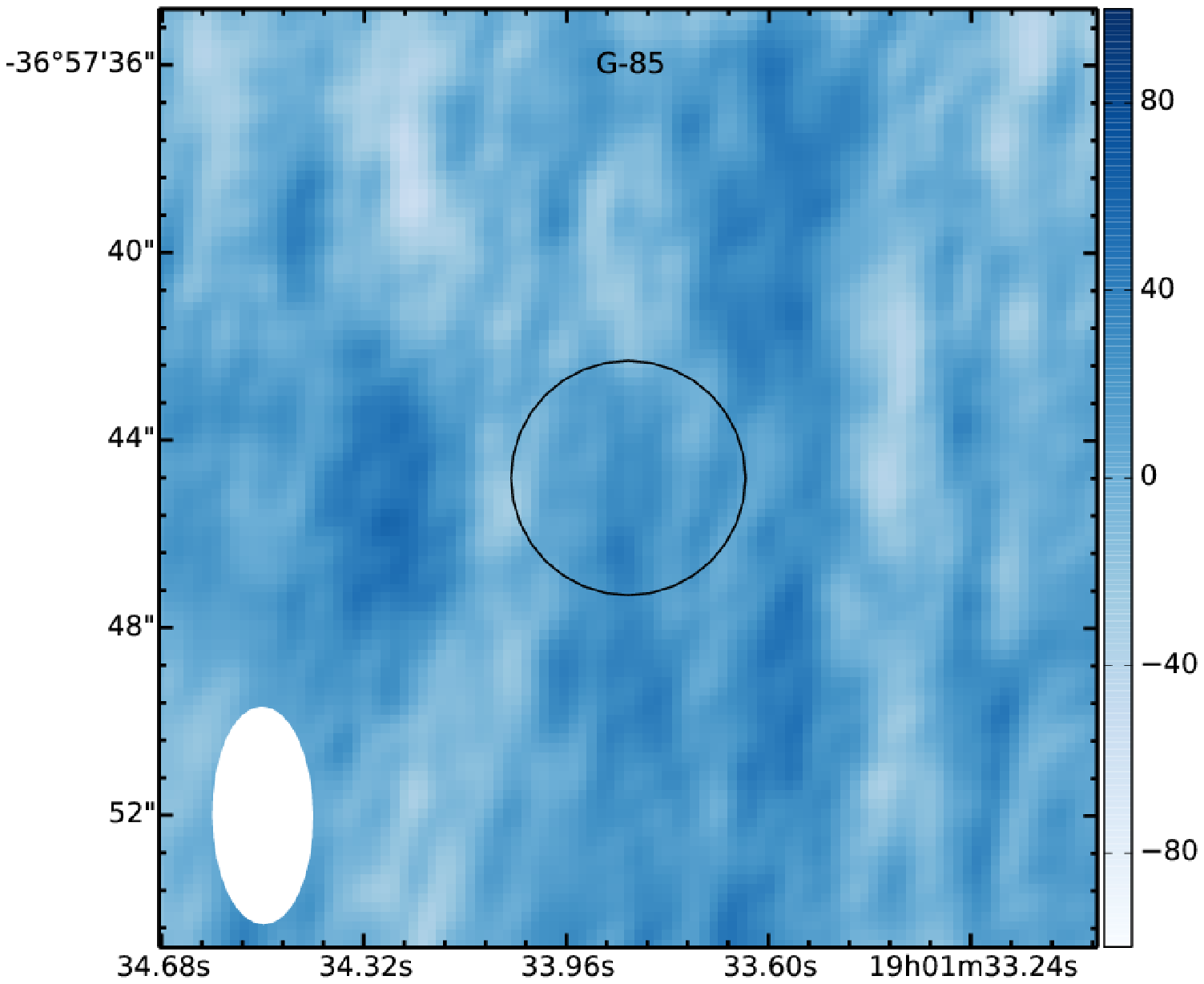} \\
\includegraphics[scale=0.25]{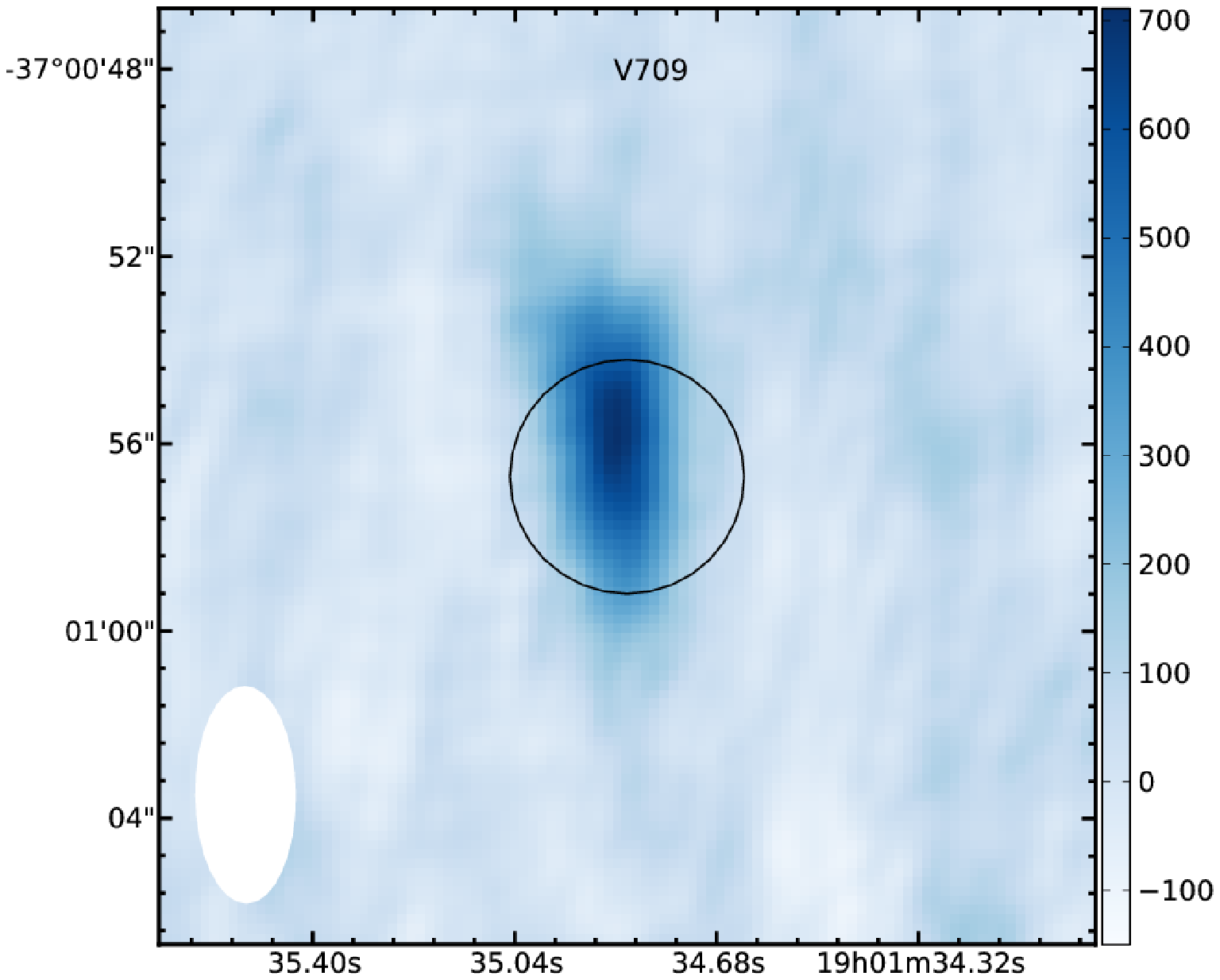} & 
\includegraphics[scale=0.25]{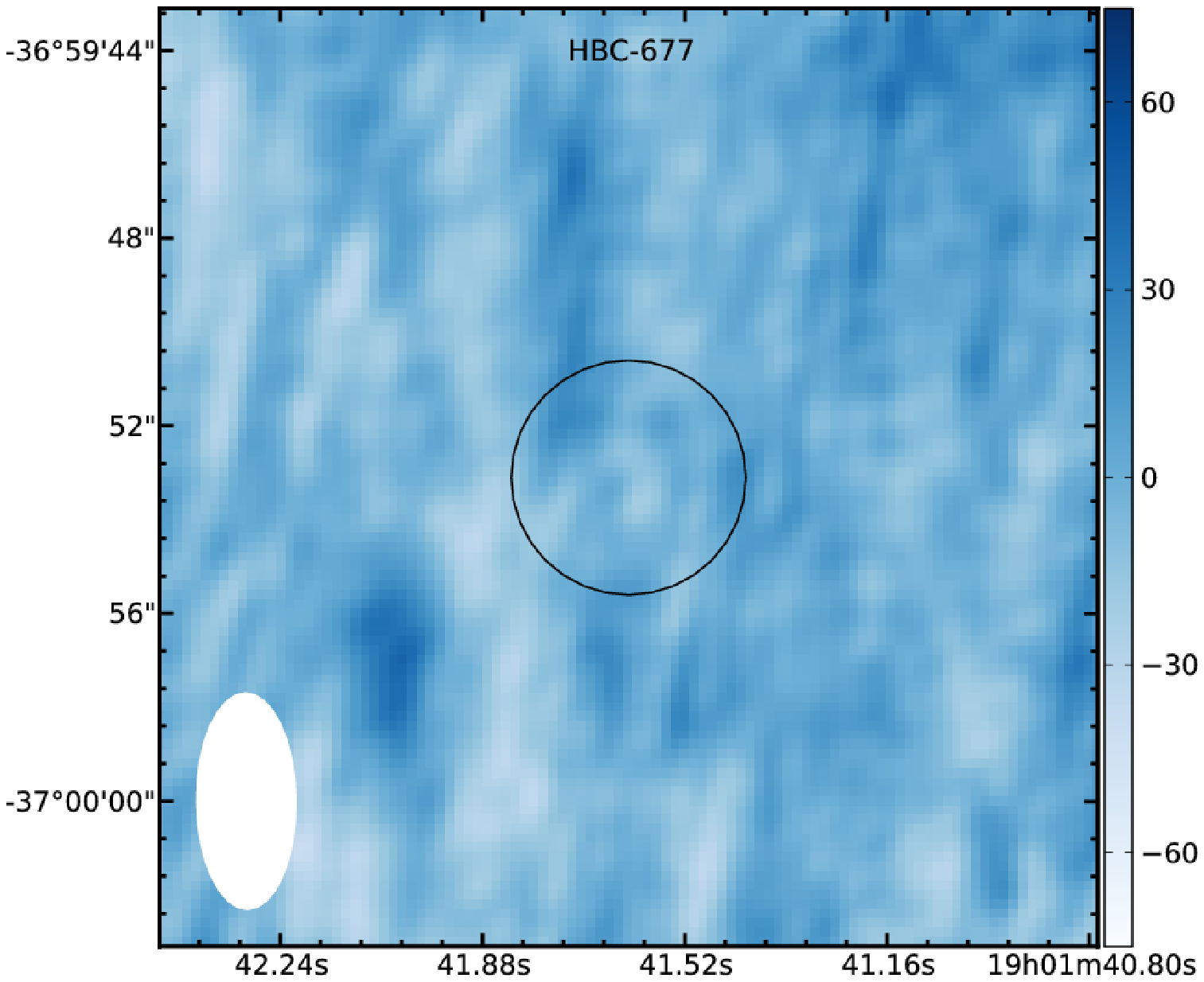} & 
\includegraphics[scale=0.25]{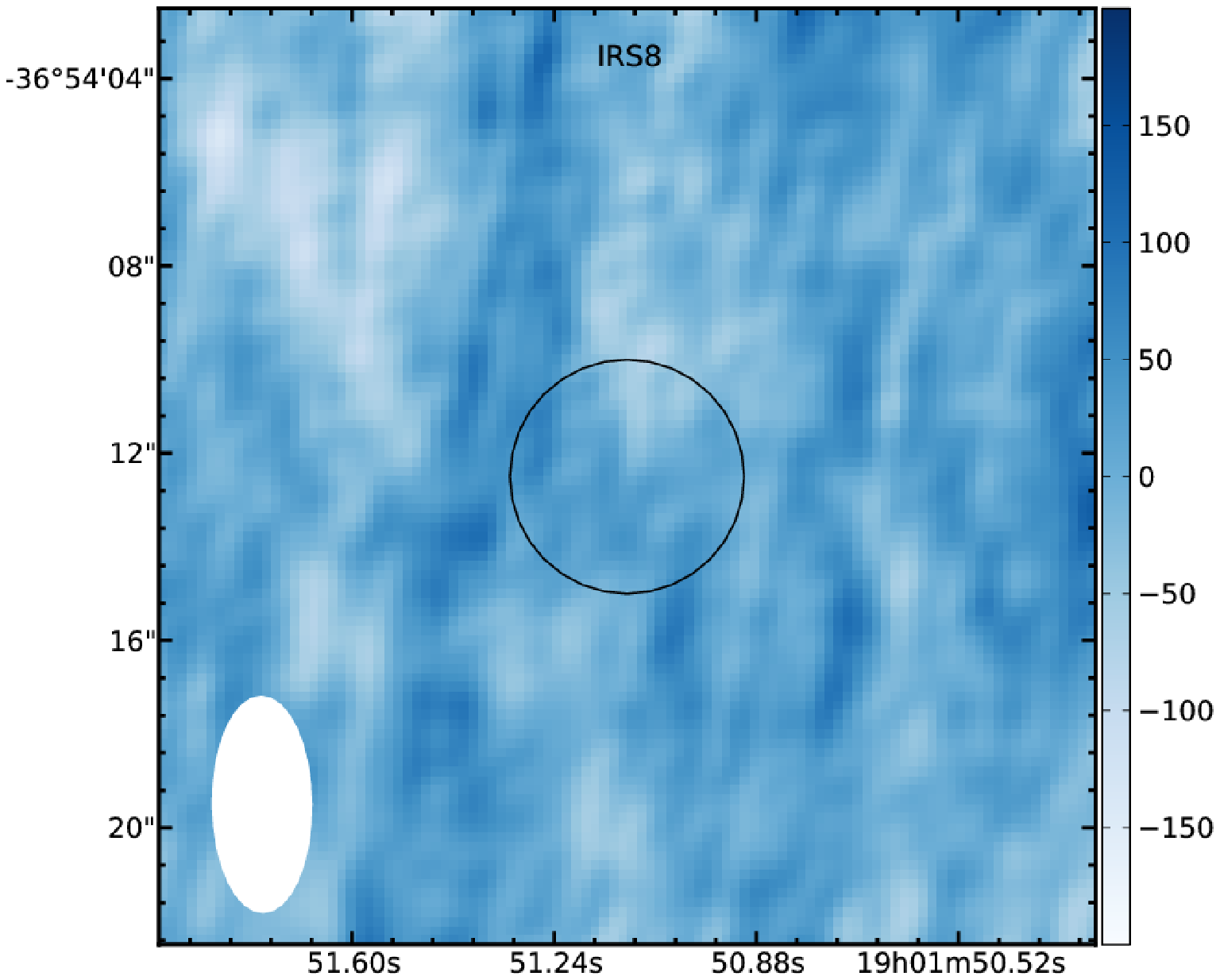} \\
\includegraphics[scale=0.25]{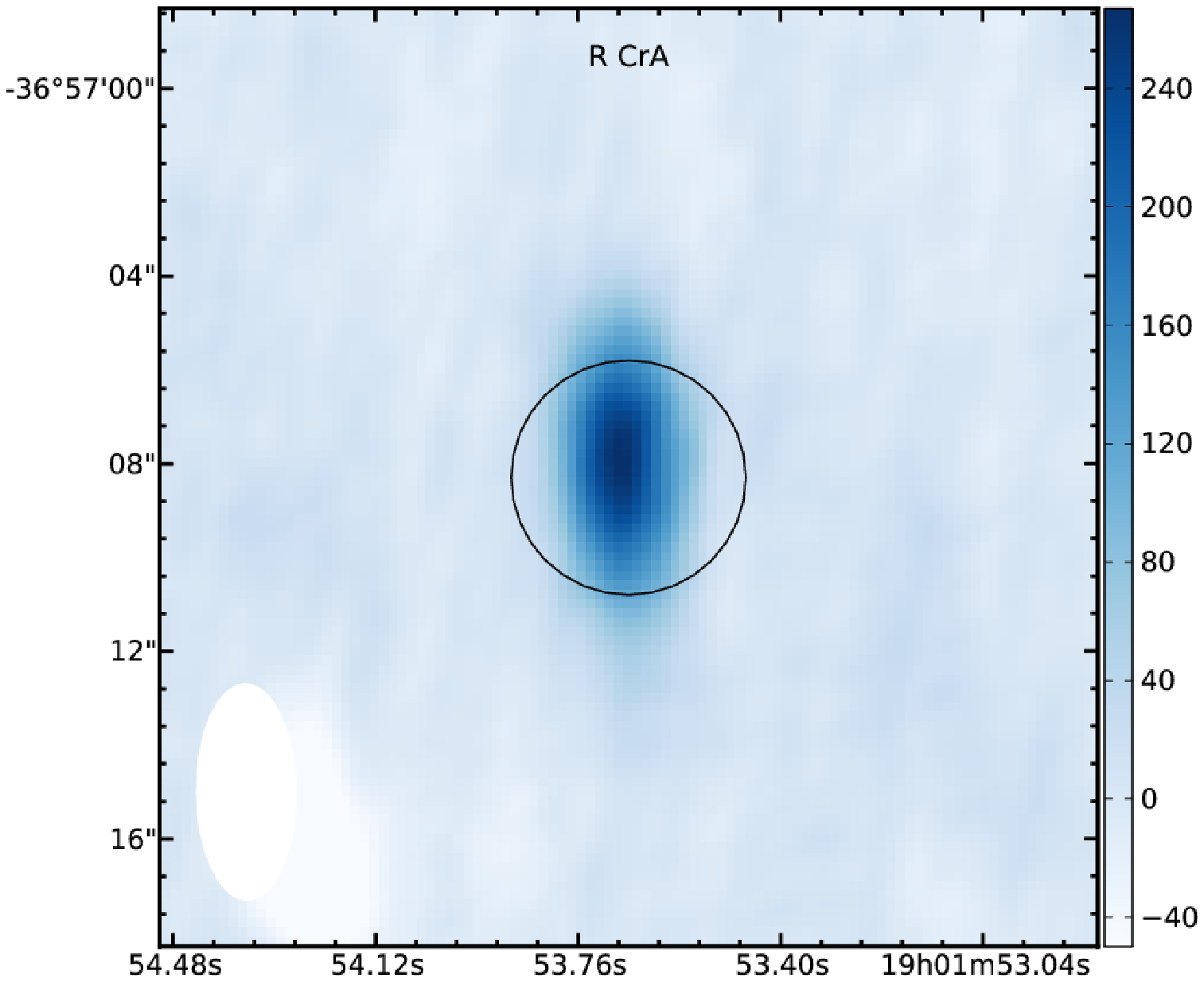} &
\includegraphics[scale=0.25]{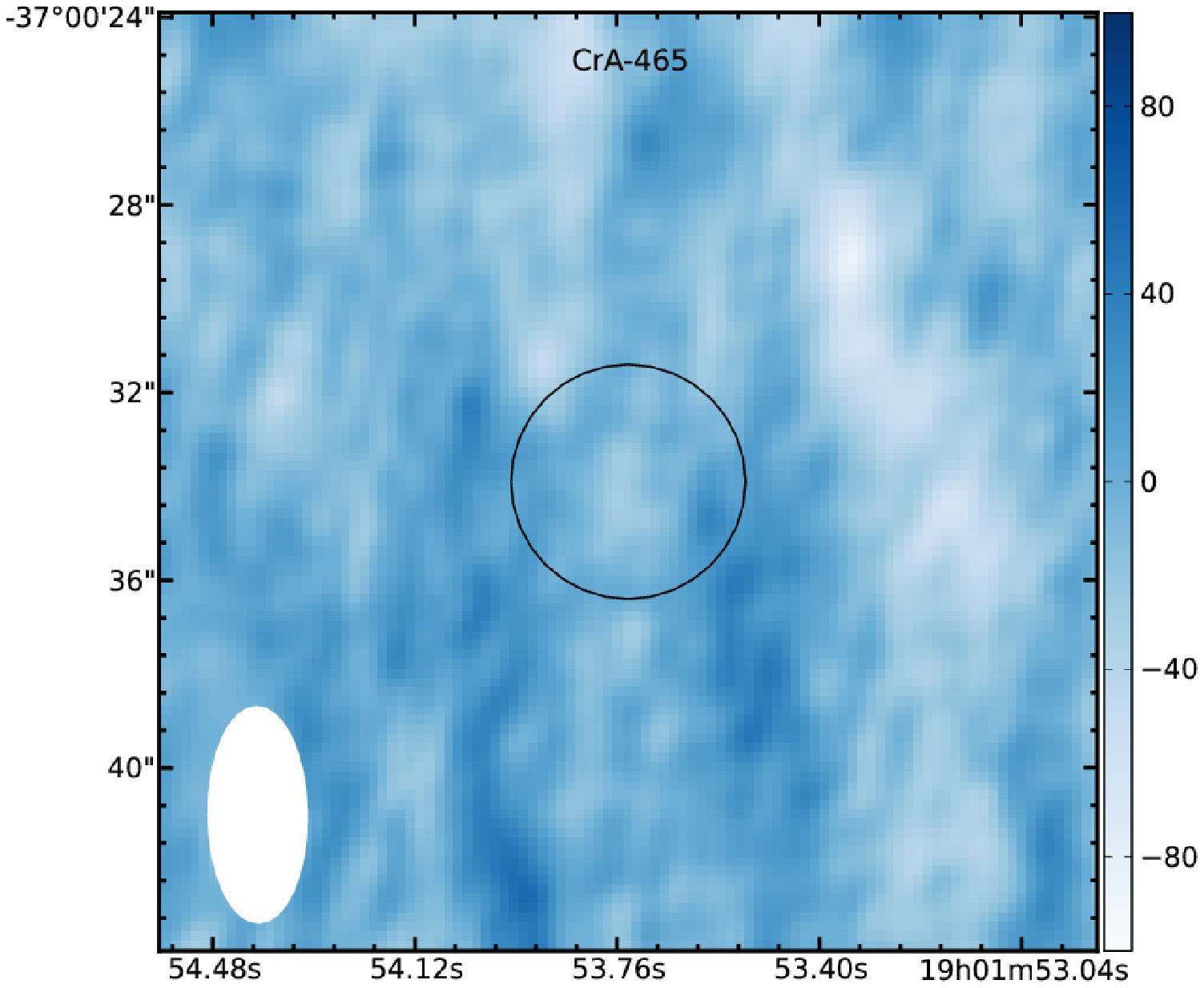} & 
\includegraphics[scale=0.25]{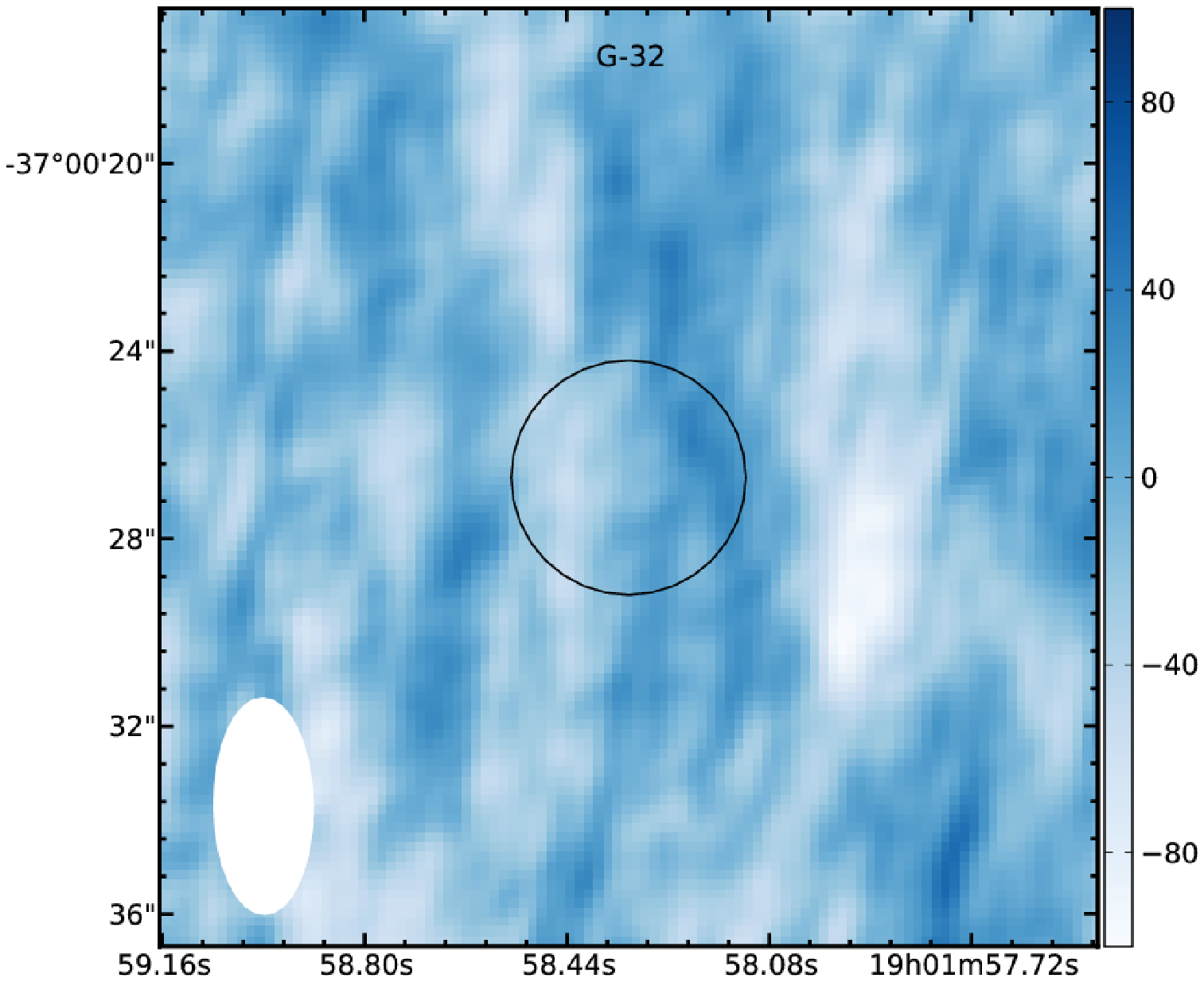} \\
\includegraphics[scale=0.25]{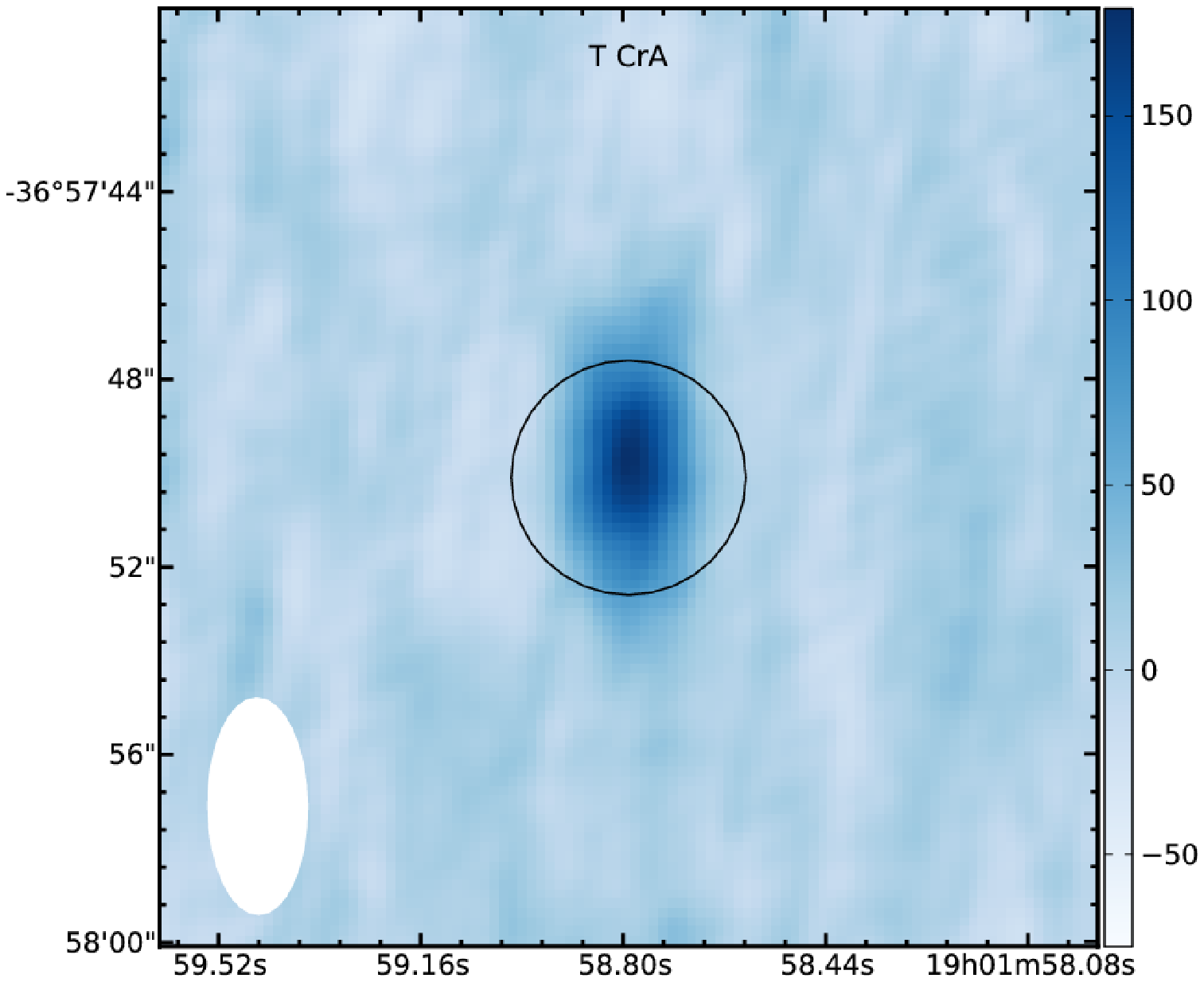} & & 
\end{tabular}
\caption{Deep VLA 3.5 cm (8.5 GHz) images of disk YSOs in CrA. 
A circle of 5\arcsec diameter is shown centered at the position given by CSA11 or SA13. 
Only the highly variable radio sources V709, R CrA, and T CrA are detected (see Table 1). 
The synthesized HPBW is $4.6\arcsec \times2.1\arcsec$, PA$=-179.4^\circ$. 
The intensity scale is in units of $\mu$Jy beam$^{-1}$.
}
\label{f2_continuum}
\end{figure}


\end{appendix}

\end{document}